\documentclass{aa}
\usepackage{graphicx}
\usepackage{epsfig}
\usepackage{txfonts}
\begin{document}
%

   \title{Speckle interferometry of the HAeBe star V376 Cas
\thanks{Based on observations performed with the 6~m telescope of the
Special Astrophysical Observatory, Russia.}}

   \author{K.W. Smith\inst{1}
          \and
	  Y.Y. Balega\inst{2}
	  \and
          K.-H. Hofmann\inst{1}
	  \and
	  Th. Preibisch\inst{1}
	 \and
	  D. Schertl\inst{1}
	 \and 
	  G. Weigelt\inst{1}
          }

   \offprints{K.Smith, \\  
              \email{kester@mpifr-bonn.mpg.de}}

   \institute{Max-Planck-Institut f\"ur Radioastronomie (MPIfR), Auf dem H\"ugel 69, 53121 Bonn, Germany 
	     \and
	     Special Astrophysical Observatory, Nizhnij Arkhyz, Zelenchuk region, Karachai-Cherkesia, 357147, Russia
             }

   \date{Received , ; accepted }

   \abstract{We report {\em H}- and {\em K}$'$-band speckle
interferometric observations of the HAeBe star V376~Cas. Our
observations show that the object is partially resolved.  The
visibility curves suggest three separate components: a large
scattering envelope visible only in the {\em H}~band, a component
approximately 100~mas in radius, and a component with a Gaussian HWHM
of approximately 8$\pm$3~mas, corresponding to approximately 5~AU at
600~pc distance, which contributes most of the flux. We compare the
smallest structure to the radius of dust sublimation in the radiation
field of the star and find that the radius is approximately six times
larger than that expected. This may indicate that the inner regions of
the system are in fact obscured by a flaring circumstellar disk or
torus seen close to edge-on.
     
   }

   \maketitle
%

\section{Introduction}

Many intermediate-mass young stellar objects (YSOs) are surrounded by
large quantities of circumstellar material, giving rise to infrared
excesses which often dominate their spectra. Much of this material
is often distributed in a circumstellar disk, through which
accretion onto the central object may continue. Whilst simple disk
models originally explained many of the features of YSOs, there
remained some outstanding problems with a thin disk
picture. Principally, for both flat passive reprocessing disks and
active accretion disks, $\lambda F_{\lambda} \propto \lambda^{-4/3}$
(Lynden-Bell \& Pringle 1974; Adams et al. 1987) whereas the
observed spectral energy distributions (SEDs) are in general flatter
than this. This implies that either a modification of the disk model
or the presence of some other component to the circumstellar material
is necessary.

Hillenbrand et al. (1992) examined the SEDs of a large sample of
HAeBe stars and divided them into three groups; those which could be
modelled with a geometrically thin disk, sometimes requiring an
optically thin inner hole, those with a rising SED at long
wavelengths, which could be modelled only with an envelope, and those
which lacked either disks or envelopes. These categories were
suggested as an evolutionary scheme for circumstellar environments of
HAeBe stars. Hartmann et al. (1993) suggested instead that
a combination of disks and envelopes was generally necessary. Since then,
disk+envelope models have been further investigated by several
authors, including for example Natta et al. (1993), Pezzuto et al. (1997), 
and Miroshnichenko et al. (1997, 1999). 

The presence of disks together with envelopes is strongly suggested by the
observations of Mannings \& Sargent (1997, 2000) who studied several
intermediate mass YSOs in the millimetre continuum and various
molecular lines.  For several sources, they found line profiles
consistent with Keplerian rotation. In some cases the rotation curve
could be spatially resolved.  The quantity of material inferred from
millimetre continuum fluxes was generally not consistent with measured
extinctions and a spherically symmetric distribution. 

The main alternative to disk+envelope models is some type of flared
disk (e.g. Chiang \& Goldreich, 1997). For a reprocessing disk, such
models modify the temperature power law and flatten the SED at
far-infrared (FIR) wavelengths. Malbet et al. (2001) and Lachaume et
al. (2003) considered flared-disk models in which the disk is heated
by both stellar light and viscous dissipation and had success in
fitting SEDs.  However, simple flared-disk models cannot reproduce the
near-infrared (NIR) `bump' between about 1 and 3~$\mu$m observed in
many HAeBe systems.  Dullemond et al. (2001) and Natta et al. (2001)
therefore proposed a new variant of the flared-disk model, introducing
an inner hole and thereby producing a bright inner rim which
contributes more NIR flux than a simple flared disk. This inner rim
can puff up and partially or completely shadow the 
more distant disk. Beyond any shadowed region, the outermost regions of the 
disk are the same as envisaged in the
simple flared-disk picture, maintaining the flux at FIR wavelengths.

In recent years, advances in speckle observing techniques, the advent
of adaptive optics, and NIR long-baseline interferometry have begun to
allow insights into the spatial distribution of the circumstellar
material and outflows associated with YSOs. For example, Leinert et
al. (2001) used a speckle technique to observe 31 HAeBe stars and
found many to have envelopes. The complex environment of the HAeBe
star R~Mon was studied by Close et al. (1997) with adaptive optics
and by Weigelt et al. (2002a) with bispectrum speckle interferometry,
revealing several structures related to the disk and outflow. Similar
bispectrum speckle observations of the high-mass YSOs S140 IRS~1 by
Schertl et al. (2000) and Weigelt et al. (2002b) and of S140 IRS~3 by
Preibisch et al. (2001) revealed complex outflows associated with
these objects.  Monin \& Bouvier (2000) used an adaptive optics system
with a resolution of about 70~mas in the {\em J} band to image an
edge-on circumstellar disk in the young triple system HV~Tau. 
Danchi et al. (2001) imaged the the massive young star MWC~349A 
using aperture-masking interferometry and 
found what appeared to be an edge-on disk. This object was also observed 
by Hofmann et al. (2002) who used bispectrum speckle interferometry and 
also saw the same probable edge-on disk.
Other edge-on disks have been imaged using HST (see for example McCaughrean
\& O'Dell 1996 or Brandner et al. 2000). Tuthill et al. (2002) used
an aperture masking interferometry technique to resolve the massive
HAeBe star LkH$\alpha$~101 and saw a circular object about 50~mas
across, corresponding to 3.4~AU at 160~pc. This object showed an
asymmetry reminiscent of an almost face-on but marginally inclined
disk. Malbet et al. (1998) resolved FU~Ori with the Palomar Testbed
Interferometer (PTI). Akeson et al. (2000) observed four YSOs with
the PTI and resolved three of them, two T~Tauri stars and an HAeBe
star.  Millan-Gabet et al. (2001) made interferometric
observations of a sample of fifteen HAeBe stars and saw many
resolved sources. Typical sizes were a few milliarcseconds,
corresponding to between about 0.5 and a few AU.  Some of the above
results were surprising, since the hot regions of classical accretion
disk models, from where the NIR flux should largely arise, should be
too small to be resolved at this level.

\begin{figure*}
  \psfig{figure=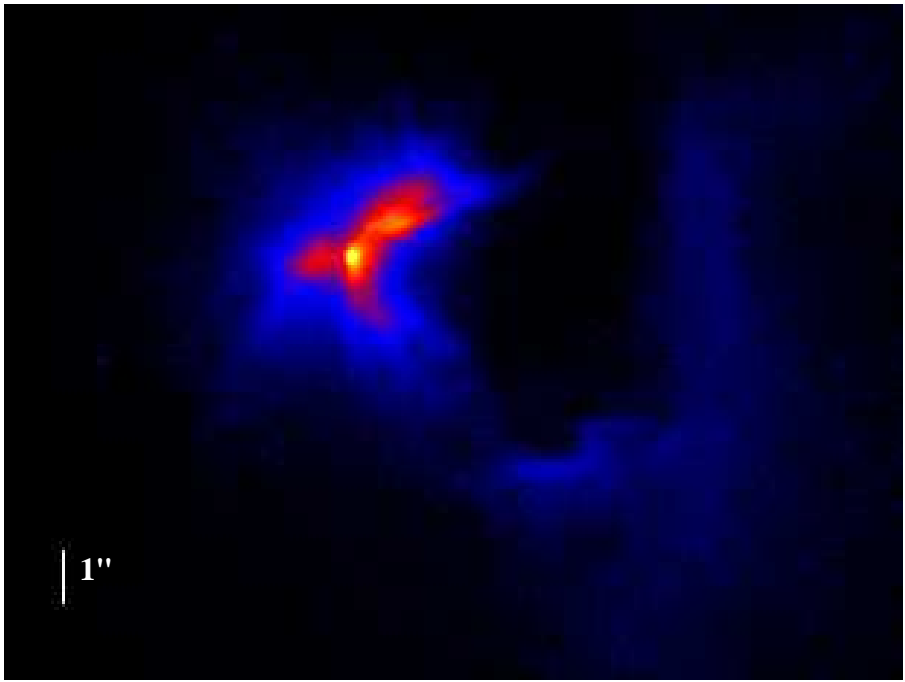,width=12.0truecm}
 \hfill      \parbox[b]{5.5cm}{\caption{ {\em HST WFPC2} image of V376~Cas taken with the F814W filter
 at 800~nm. This image, which was obtained from the archive, is the
 combination of one 100~s and one 300~s exposure, both of which were
 made in March 2000. North is up and East is to the left. 
The brightest knot probably corresponds to
the component observed at {\em H} and {\em K}$'$. }}
          \label{hst}%
     \end{figure*}

V376~Cas is a HAeBe star embedded in the small dark cloud L1265 and
associated with considerable nebulosity. At optical wavelengths, on
arcsecond scales, this nebulosity is seen to be elongated along an
axis with position angle approximately 120$^{\circ}$ (Leinert et
al. 1991; Piirola et al. 1992; Corcoran et al. 1995; Asselin et
al. 1996).  The source displays an extreme degree of linear
polarization in the red, $\sim 20 - 25$\%, higher than for any other
intermediate mass YSO, indicating that a high proportion of the
emerging light has been scattered. Polarization maps were produced at
900~nm by Leinert et al. (1991), in the {\em I} band by Piirola et
al. (1992), and in the optical by Asselin et al. (1996). These maps
show strong centrosymmetric polarization in the lobes of the elongated
nebulosity and areas of low polarization situated near the star on
either side of the extended envelope. Leinert et al. (1991) used
speckle polarimetry at {\em K} to show that this pattern persists to
subarcsecond scales. This pattern of polarization suggests
a bipolar reflection nebula aligned
northwest-southeast, which in turn implies the presence of an edge-on
disk at a position angle of approximately 30$^{\circ}$.  
{\em HST~WFPC2} images obtained from the archive (Fig.~\ref{hst}) confirm
this elongated source. However, at this higher resolution, the overall
nebulosity is not reminiscent of an edge-on disk with position angle
30$^{\circ}$, but rather suggests an outflow cavity to the west of the
object. This in turn suggests a nearly edge-on disk at
position angle approximately 140$^{\circ}$, together with the nearside
lobe of a bipolar outflow extending to the west. The observed light
probably arises mostly from scattering from the inner edge of the
outflow cavity or from the surface of a flared disk or circumstellar
torus.

\section{Observations and data reduction}

The speckle interferograms were recorded in September 2002 with the
SAO 6~m telescope in Russia. The detector of the speckle camera was a
Rockwell HAWAII array. The two filters used corresponded approximately
to {\em H} (central wavelength 1.648~$\mu$m, bandwidth 0.317~$\mu$m)
and {\em K}$'$ (central wavelength 2.115~$\mu$m, bandwidth 0.214~$\mu$m).
Further observational parameters are listed in Table 1.

\begin{table*}
\caption{Observational parameters.  $\lambda_c$ and $\Delta\lambda$
give the central wavelength and bandwidth of the filter. FoV is the
field of view of both the data recording and data processing. $N_{\rm
T}$ and $N_{\rm R}$ are the total numbers of speckle interferograms of
V376~Cas and the reference star, respectively, and $N_{\rm P}$ is the
number of target-object pairs observed per night.  T is the exposure
time per frame and S is the Seeing (FWHM). In the last column, the
names of the reference stars are given.  }
\begin{center}
\begin{tabular}{llllllllll}
\hline 
\hline 
Observation Day & $\lambda_c$ & $\Delta\lambda$ & FoV    &   $N_{\rm T}$ & $N_{\rm R}$ & $N_{\rm P}$ & $T$\hspace*{1.2mm}&  Seeing              & Reference \\
Sep 2002        & [nm]        & [nm]            & ($''$) &               &             &             &  [ms]             & [$^{\prime\prime}$]  & Star \\
\hline
20              & 2115        & 214             & 5.2  & 562         & 732         & 1 & 164            & 2.2                  & HD 236352 \\
22              & 2115        & 214             & 5.2  & 275         & 536         & 1 & 245            & 3.2                  & HD 236352 \\
23              & 2115        & 214             & 5.2  & 368         & 750         & 1 & 164            & 2.2                  & HD 236352 \\
25              & 2115        & 214             & 5.2  & 635         & 1264        & 2 & 184            & 2.1                  & HD 236324 \\
\hline
24              & 1648        & 317             & 3.8  & 2083        & 3160        & 5 & 180            & 1.7                  & HD 236324 \\
\hline
\label{obstable}
\end{tabular}
\end{center}
\end{table*}

The modulus of the object Fourier transform (visibility) was obtained
with the speckle interferometry method (Labeyrie 1970). Speckle
interferograms of calibrator stars, which were measured to be unresolved,
were recorded just before
and just after the object and served as reference stars for the
determination of the speckle transfer function (see Table~\ref{obstable}).

\begin{figure*}
\caption{ Upper panels: Two-dimensional visibilities of V376~Cas. Left column
filter 2115/214\,nm.  Right column filter 1648/317\,nm.  The contours are 
drawn at levels 0.95, 0.90, 0.85, 0.80, 0.75 and 0.70.
Lower panels:
Azimuthally averaged visibilites.  Left column filter 2115/214\,nm.
Right column filter 1648/317\,nm.  The individual dots show the
observed visibilities, the solid line the fitted model consisting of
two Gaussians at 2115\,nm and three Gaussians at 1648\,nm.}  
\vbox{ 
\hbox{
\epsfig{figure=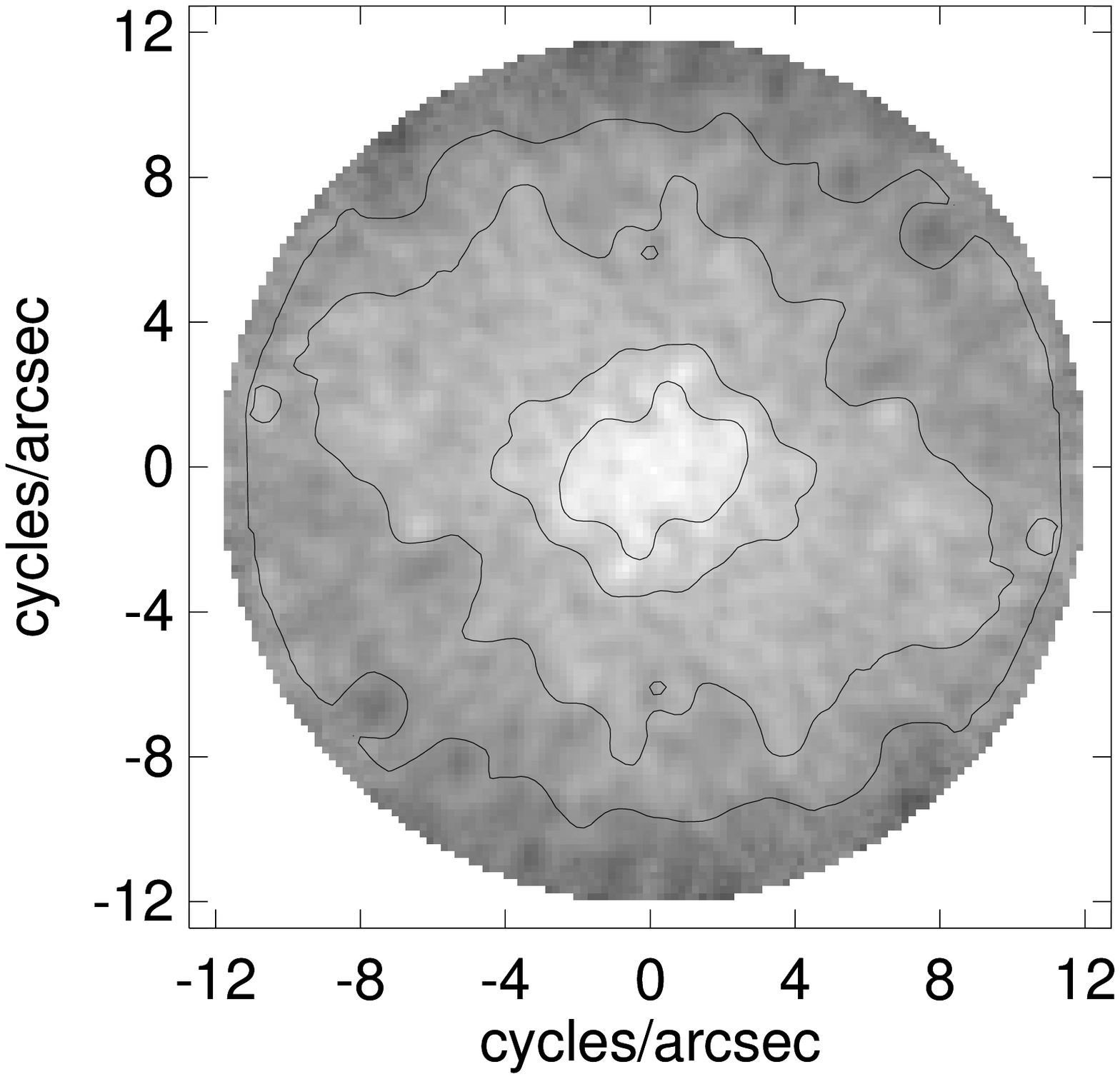,width=8.truecm,height=8.truecm}
\epsfig{figure=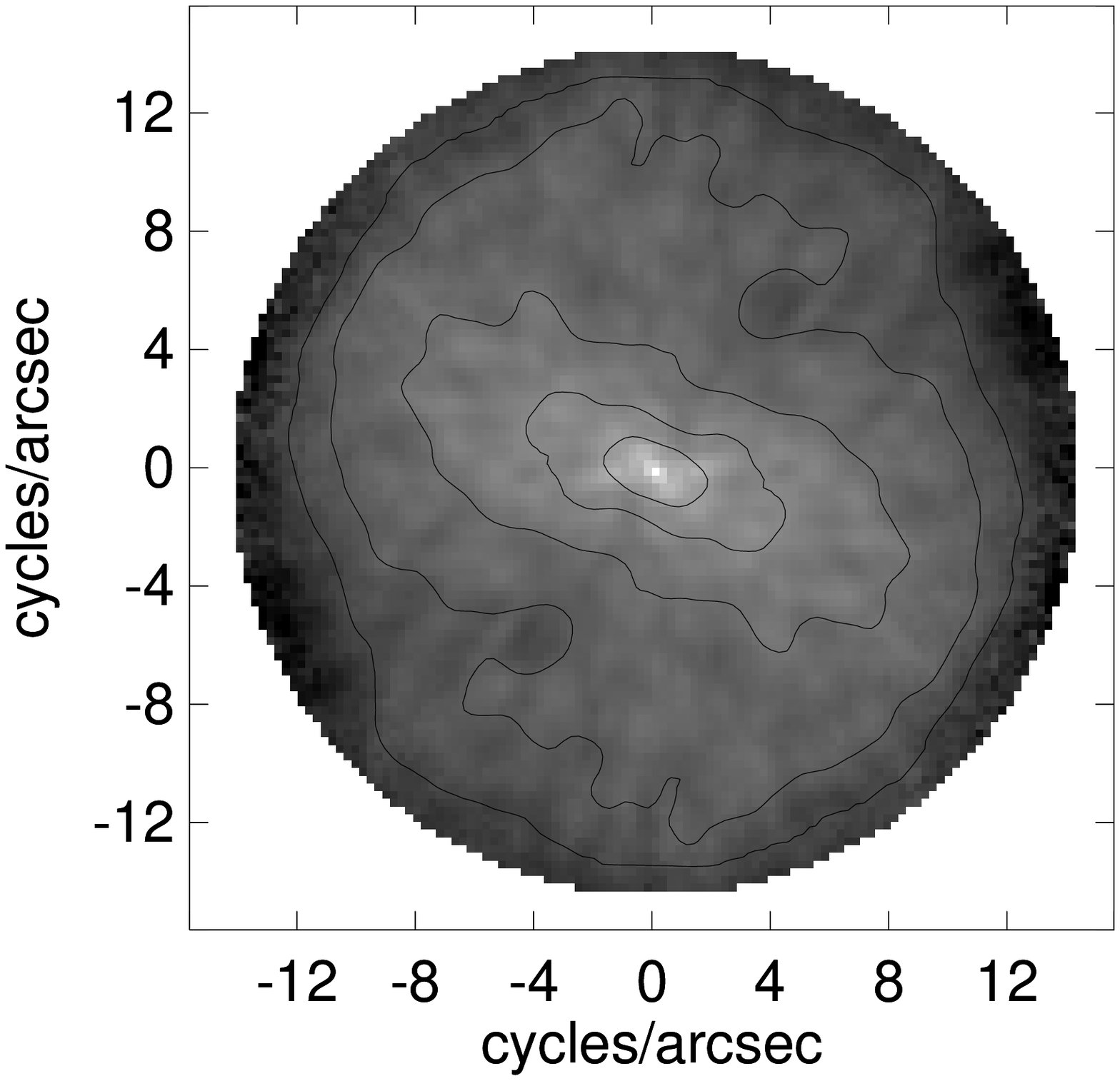,width=8.truecm,height=8.truecm} } \hbox{
\epsfig{figure=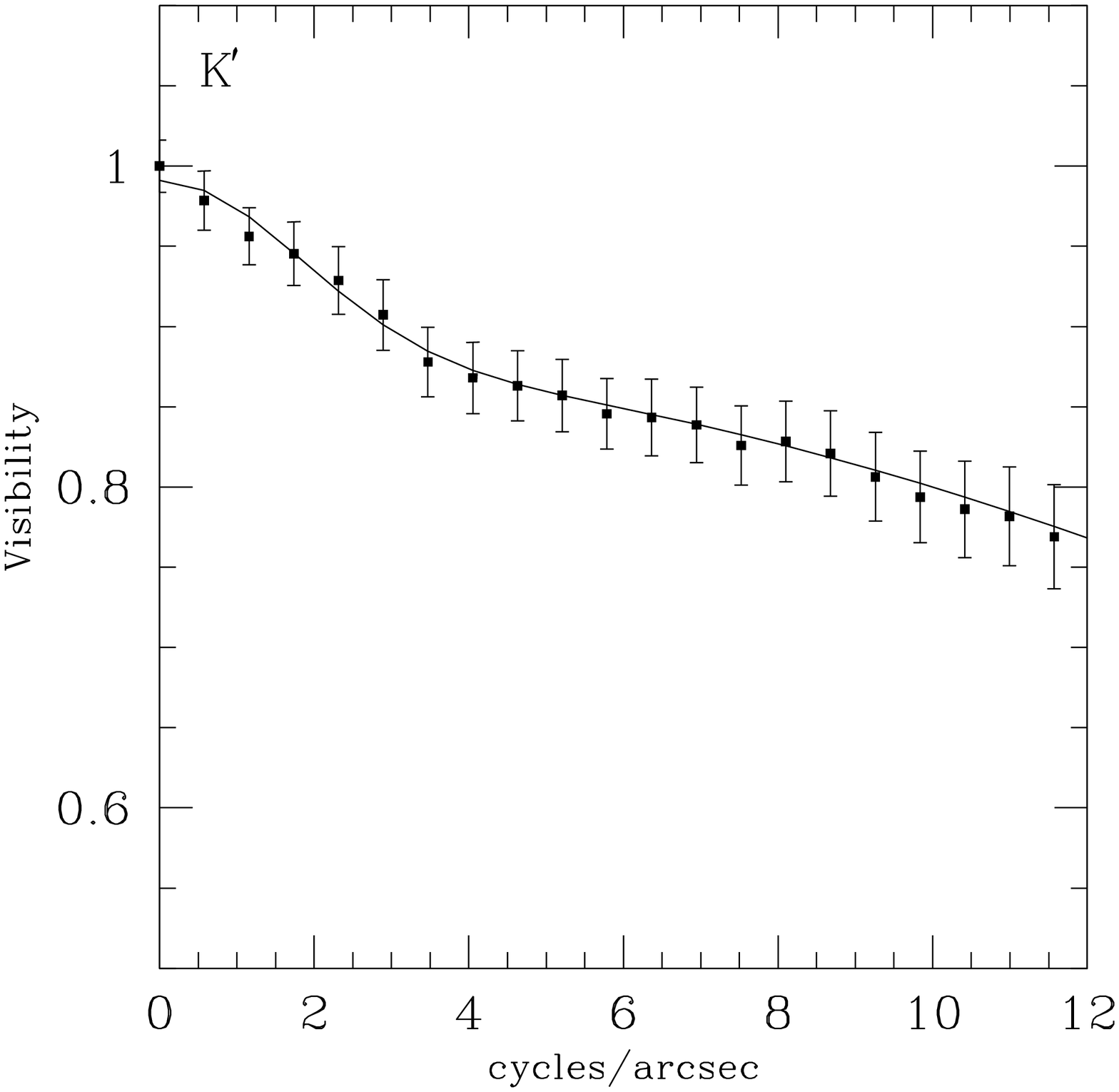,width=8.truecm,height=8.truecm}
\epsfig{figure=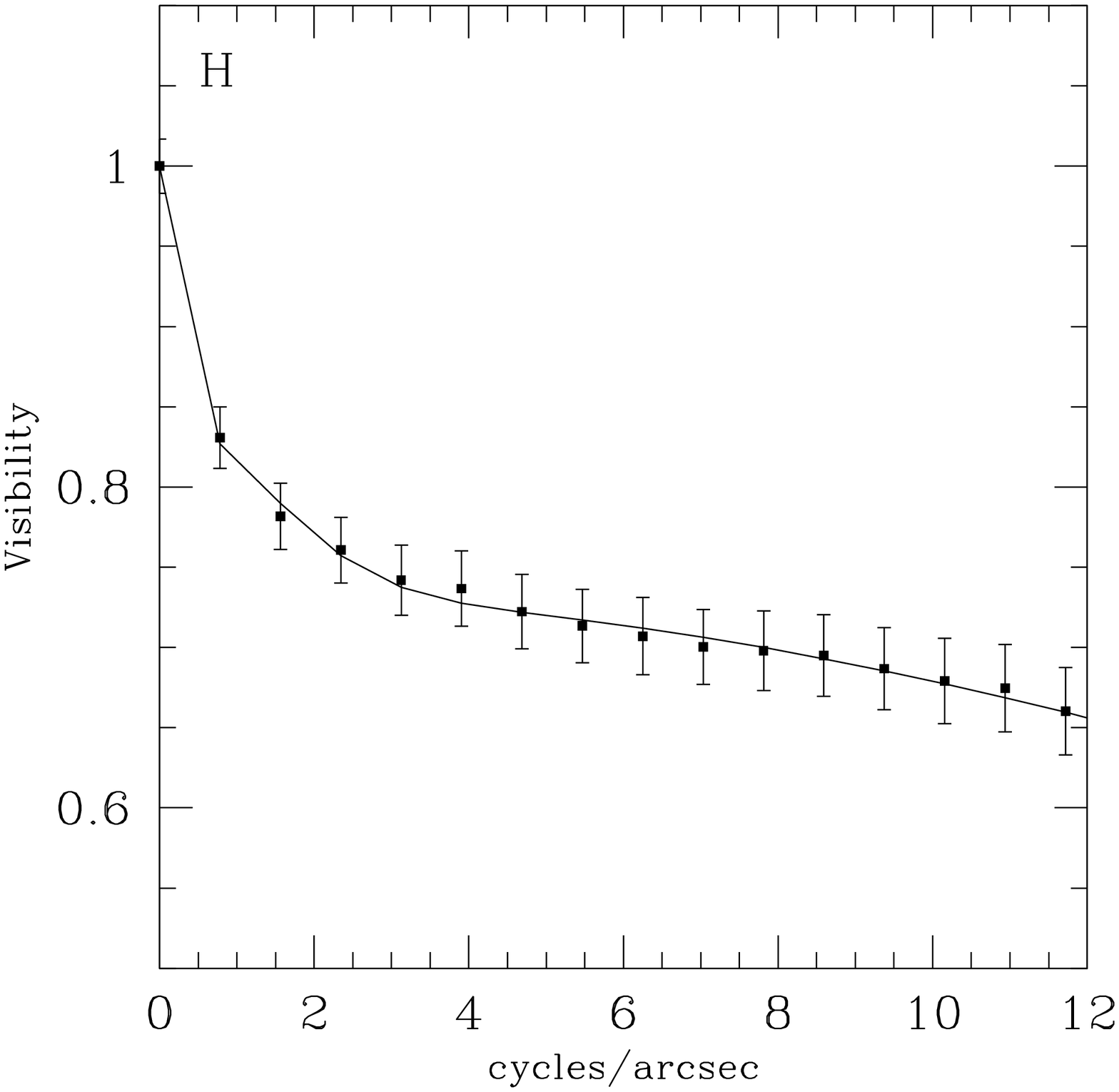,width=8.truecm,height=8.truecm} 
}}
\label{visplots}
\end{figure*}

\section{Results and discussion}
\label{results}

The reconstructed visibilities of V376~Cas are presented in
Fig.~\ref{visplots}. The upper two panels show the two-dimensional
visibilities in {\em K}$'$ (left) and {\em H} (right).  The
visibilities show some evidence of elongation along a position angle
approximately 70$^{\circ}$ from the vertical. The {\em H}-band visibility is
dominated by a central spike, indicating the extended emission in the
3.8$''$ field of view of {\em H}, which is not evident at {\em K}$'$.
The azimuthally averaged visibilities are shown in the two lower
panels of Fig.~\ref{visplots}. These plots have both been
truncated at twelve cycles per arcsecond, beyond which the quality of
the data is significantly degraded by noise. There is clear evidence
of resolution at both wavelengths. The {\em K}$'$-band visibilities
decline by about 20\% and the {\em H}-band visibilities by about 30\%
over the range of spatial frequencies sampled.  Both curves show a
significant decrease in steepness at $\sim$4 cycles/arcsec. This
structure suggests a two-component visibility model for the {\em K}$'$
band and a three-component visibility model for the {\em H} band.  The
best-fitting 3- and 2-component Gaussian models for {\em H} and {\em
K}$'$ are shown in the plots.  A second model was also fitted, in
which the most compact Gaussian component (on the sky) is replaced by
a uniform-brightness ring.  The inner radius of the ring was the
fitted parameter, and the outer radius was set at 110\% of this.  The
intention of fitting this model is to allow direct comparison with the
results reported for similar objects in the literature
(e.g. Millan-Gabet et al. 2001).  To determine uncertainties in the
fitted parameters, each individual target-reference star pair was
fitted separately and the resulting parameters averaged. There were
five such independent observations in total in each band. The
peak-to-peak variation of each parameter was then used as an estimate
of the uncertainty. Table~\ref{params} lists the parameters and
associated uncertainties of all these fits.

The Gaussian fit components are an approximately 1$''$ HWHM extended
nebula visible only in {\em H}, an approximately 100~mas HWHM
component, corresponding to about 70~AU in radius at an assumed
distance of 600~pc (Chavarria-K 1985), and a small, approximately
8$\pm 3$~mas HWHM component comprising a large fraction of 
the flux in each band,
corresponding to a region of about 4.8~AU in radius at 600~pc (see
Fig.~\ref{visplots}).  It should be stressed that smaller
components could in fact be present which are not resolved at our
longest baselines. We cannot, of course, exclude the possibility of
another break in the visibility curve at baselines longer than 6~m.

The H-band visibility function shows resolved structure both
along and across the elongation. This, together with the approximately
circularly symmetric appearance of the K$'$-band visibility function
allows us to rule out a binary model as the sole explanation for the
resolved source and to identify at least part of the resolved
structure with circumstellar material. It is possible that 
a companion is also present, with a position angle of approximately 
160$^{\circ}$ or 340$^{\circ}$, and contributes partially 
to the resolved structure. It is also possible that the
elongation is due to the presence of a nearly edge-on disk. 
If this were so, the disk plane would run along a direction perpendicular to 
the observed elongation in the visibility function, that is, from northwest to
southeast. Such a disk would be broadly compatible with the interpretation
offered above for the HST image, showing an outflow cavity extending 
to the southwest.

\subsection{The 8~mas component: Dust sublimation?}

An explanation often found in the literature for the observed NIR sizes
of YSOs is that the observed structure lies at the dust sublimation radius, 
that is, the radius at which circumstellar dust sublimates due to the
radiation field of the central star. This gives rise either 
to a hot inner edge of a disk, as in the model of Dullemond et al.
and similar models, or to a sharp inner boundary to a spherically symmetric 
shell. This type of explanation was first suggested for individual objects by 
Tuthill et al. (2001) and Natta et al. (2001), and was investigated 
for a sample of resolved YSOs by Monnier \& Millan-Gabet (2002).
Monnier \& Millan-Gabet (2002) demonstrated that the NIR
sizes of various HAeBe and T Tauri stars appearing in the literature
correlates with the stellar luminosity in a way that is consistent with their
being determined by dust sublimation. Here, we compare 
the smallest resolved component of V376~Cas with the 
expected dust sublimation radius for this object.

The dust sublimation radius can be estimated as 
\begin{equation}
\begin{array}{lll}
R_{s} & = & \frac{1}{2} \sqrt{Q_R}\left(\frac{T_*}{T_s}\right)^2 R_* \\
      & = & 1.1 \sqrt{Q_R} \left(\frac{L_*}{1000L_{\odot}}\right)^{1/2}
            \left(\frac{T_s}{1500 K}\right)^{-2} \mathrm{AU},
\end{array}
\label{dustsub}
\end{equation}
where $Q_R$ is the ratio of the incident absorption efficiency to the
re-emitted absorption efficiency (Monnier \& Millan-Gabet 2002).  

For hot stars, $Q_R$ can vary strongly if the grain size is smaller
than about one micron. For grains of 1~$\mu$m or larger, $Q_R \sim 1$,
but $Q_R$ may reach values of around 50 or more for grain sizes of
less than 0.25~$\mu$m and stellar surface temperatures of 10,000~K or
greater.  However, if large grains existed in sufficient quantities, they
would survive longer and determine the position of $R_s$. This logic
was followed by both Monnier \& Millan-Gabet (2002) and Tuthill et al. (2001), who
both assumed $Q_R=1$.

\begin{figure}
\psfig{figure=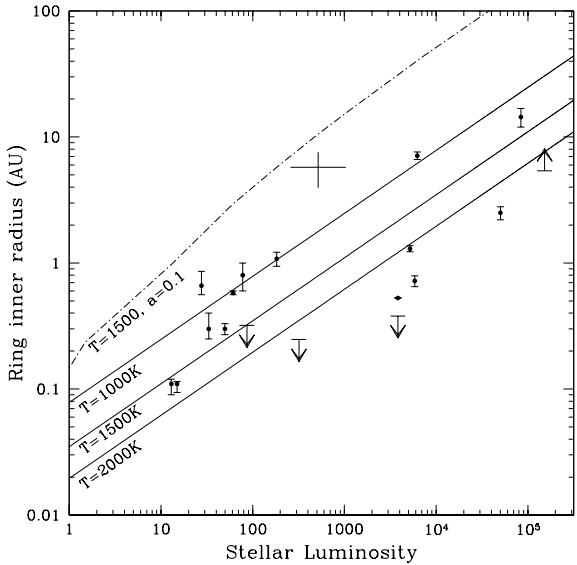,width=8.truecm,height=8.truecm}
\caption{The sizes of YSOs as a function of stellar luminosity, based
on Fig.~1 of Monnier \& Millan-Gabet (2002). The data points show
the inner ring radii of uniform-ring fits to resolved sources from
Millan-Gabet et al. (2001), the measured size of LkH$\alpha$~101 from
Tuthill et al. (2002), and the HWHM for Gaussian fits to two classical
T~Tauri stars from Akeson et al. (2000).  The cross marks the position
of V376~Cas.  The luminosity has been taken as 517~L$_{\odot}$ with an
assumed error of a factor of two. The radius plotted is the inner ring radius 
of the smallest component fitted to the {\em K}$'$ data. The
straight lines mark the model with $Q_R=1$ for three dust sublimation
temperatures, 1000~K, 1500~K, and 2000~K. The dashed line is a model
with a grain size of 0.1~$\mu$m and $T_s=1500$~K.}
\label{radiuslum}
\end{figure}

The maximum temperature at which dust grains can survive is estimated
to be around 1750~K for graphite and about 1400~K for silicate grains
(Hillenbrand et al. 1992). We use a sublimation temperature of
1500~K but also consider temperatures of 1000~K and 2000~K.

\begin{table*}
\begin{center}
\caption{Results of circular symmetric fits to the observed
visibilities. Two different models were fitted at each wavelength.
The first model consisted of either two (at {\em K}) or three (at {\em
H}) Gaussians. For the second model, the broadest Gaussian in the
visibilities (corresponding to the narrowest component on the sky) was
replaced with a uniform-brightness ring, the inner radius of which was
a fitted parameter and the outer radius of which was set to 110\% of
the inner radius.}
\label{params}
\begin{tabular}{lllllll}
\hline 
\hline 
        \multicolumn{7}{c}{Multiple Gaussian models} \\ 
                                                     \\
\hline
        & \multicolumn{2}{c}{Small Component} & \multicolumn{2}{c}{Medium Component} & \multicolumn{2}{c}{Large Component} \\
Filter  & HWHM        & Relative              & HWHM         & Relative              & HWHM       & Relative            \\
 $[nm]$ &      [mas]  &      flux             &      [mas]   &      flux             &      [mas] &      flux           \\
\hline
2115   & 8.2$\pm 3$   & 88$\pm 4$\%          & 105$\pm 50 $  & 11$\pm 5$ \%   & ---                     & ---            \\
1648   & 7.4$\pm 3$   & 73$\pm 3$\%          & 134$\pm 50 $  & 11$\pm 5$ \%   & 711$^{+700}_{-350}$  & 16$\pm 4$ \%    \\
\hline \\
         \multicolumn{7}{c}{Uniform-brightness ring plus Gaussian models} \\
\\
\hline
        & \multicolumn{2}{c}{Small Component} & \multicolumn{2}{c}{Medium Component} & \multicolumn{2}{c}{Large Component} \\
Filter  & Ring inner        & Relative        & HWHM         & Relative              & HWHM           & Relative            \\
 $[nm]$ &  radius [mas]     &  flux           & [mas]        &      flux             &      [mas]     &  flux           \\
\hline       
2115   & 9.6$\pm 3$         & 88$\pm 4$\%           & 104$\pm 50$ & 10.4$\pm 4$\%               & ---            & ---             \\   
1648   & 8.3$\pm 3$         & 73$\pm 4$\%           & 124$\pm 80$ & 10.9$\pm 4$\%               & 843$\pm 700$  & 16.6$\pm 3$\%        \\ 
\hline
\end{tabular}
\end{center}
\end{table*}

Various estimates of the luminosity of V376~Cas are available in the
literature.  Chavarria-K. (1985) estimated the luminosity of V376~Cas
to be 517~$L_{\odot}$, based on his own extinction and distance
estimate and the photometry of Allen (1972) and Cohen
(1974). \'Abrah\'am et al. (2000) observed V376 in the mid to far
infrared with ISOPHOT and derived a luminosity between 1 and 1000 
microns of 430~$L_{\odot}$. Since the vast majority of the flux
emerges in this wavelength range, this value is probably not a serious
underestimate due to the wavelength truncation at either end. Natta
et al. (1992) estimated the luminosity of V376~Cas to be about
150~$L_{\odot}$, based on a combination of new FIR observations and
existing optical and NIR measurements. For the purposes of this
analysis, we adopt the luminosity of 517~L$_{\odot}$ given by
\'Abrah\'am et al. (2000), together with an uncertainty of a factor of
two.

With these assumptions ($L_*=517~L_{\odot},\ T_s=1500\ \mathrm{K},\ 
Q_R=1$) the dust sublimation radius for V376~Cas given by
Equation~\ref{dustsub} is 0.79~AU.  This rises to 1.78~AU if the dust
is destroyed at a temperature of 1000~K. These values lie
approximately a factor of three to six below the HWHM of the smallest
fitted Gaussian for V376~Cas, at about 4.8~AU.  This is illustrated in
Fig.~\ref{radiuslum}, which is based on Fig.~1 of Monnier \&
Millan-Gabet (2002).  This plot shows the relationship between stellar
luminosity and dust sublimation radius for dust grains of 1~$\mu$m or
larger and three different dust sublimation temperatures, as well as the
dust sublimation radius for a dust destruction temperature of 1500~K
and a grain size of 0.1~$\mu$m. The position of V376~Cas in the {\em
K}$'$ band is marked.  The error bars in the size are determined from
the fitting procedure, as described in Sect.~\ref{results}
above. The luminosity is assumed to be uncertain by a factor of
two. No account is taken of possible errors in the distance of
600~pc. This is because an error in the distance would move the point
parallel to the model lines in the figure and would therefore have no
effect on any comparison between the data point and the models.  Also
shown are points for a number of other HAeBe and T~Tauri stars taken
from the literature (see caption for references).

\subsubsection{Comparison with models}

The most compact component of V376~Cas which we resolve is about six times larger than
the basic dust sublimation model predicts and also appears large
compared to the population of similar objects shown in
Fig.~\ref{radiuslum}. Two possible explanations for this are that
the dust grain population is dominated by grains significantly smaller
than 1~$\mu$m, and that the inner regions of the system are
obscured by a disk or torus viewed close to edge-on.

In Fig.~\ref{radiuslum} we show the estimated dust sublimation
radius for grains of 0.1~$\mu$m in size. Such a grain size leads to a
predicted dust sublimation radius significantly larger than
the measured size of V376~Cas, demonstrating that even a modest change
in grain properties might accommodate our result.  The scattering
envelope seen at {\em H} but not at {\em K}$'$ may indicate that the
scattering cross section is very much smaller at {\em K}$'$ than at {\em
H}. This could imply that the population of dust grains in this
extended region is dominated by small grains.  However, the position
of V376~Cas in Fig.~\ref{radiuslum} is not only somewhat high
compared to the models but also compared to other similar
objects. Invoking a small dust-grain size to explain this would
therefore involve suggesting that V376~Cas is in this respect an
exceptional case.

Can the luminosity estimate for V376~Cas be in error?  The spectral
type of V376~Cas was determined by Cohen \& Kuhi (1979) from low
resolution spectra to be B5. Earlier, Herbig (1960) estimated the
spectral type of V376~Cas to be late B or early A.  Rodriguez \& Canto
(1983) observed the field at 6~cm and saw no corresponding radio
source down to a level of approximately 1~mJy, which implies that the
spectral type cannot be much earlier than mid B.  Comparison with
evolutionary tracks (e.g. Palla \& Stahler 1993; Driebe, 2001)
suggests that a B5 star of surface temperature 14,000~K should have
luminosity of between approximately 200 and 1000~$L_{\odot}$. This is
broadly consistent with the Chavarria-K. (1985) and \'Abrah\'am et
al. (2000) estimates. However, if the source has a disk, the escaping
radiation field will be anisotropic, and if the disk is viewed close
to edge-on, this would lead to the photometrically determined
luminosity, and hence the expected dust sublimation radius, being
{\em underestimated}, probably by a factor of approximately two to
four. Such an underestimate is easily compatible with a small
uncertainty in the spectral type.

Finally, if the NIR flux arises mostly at the inner edge of a
disk, as in the model of Lachaume et al. (2003), Dullemond et
al. (2001), and Natta et al. (2001), this region might well be
obscured from view by the outer parts of any flaring disk or
circumstellar torus.  The observed flux would then be reflected from
the circumstellar material, and the size of the dominant component in
the NIR might well be considerably larger than the original dust
sublimation radius. This possibility is of course entirely consistent
with the high level of linear polarization observed from the object.

\subsection{The other components}

The largest structure seen at {\em H} but not at {\em K}$'$ clearly
corresponds to the structure seen by Leinert et al. (1991) and also by
Piirola et al. (1992) and others. This extended {\em H}-band component is
caused by scattered light from the extended nebula. This intermediate
structure (200~mas $\sim$ 120~AU or 60~AU in radius) may represent
scattering from the flaring parts of the outer disk and probably
corresponds to the bright knot seen in the {\em HST } images.

\section{Conclusions}

Our {\em H}- and {\em K}$'$-band speckle interferometric observations
of V376~Cas suggest two resolved compact components, with
characteristic radii of 100~mas and 8~mas. Additionally, a scattering
envelope about 2$''$ in diameter is seen in the {\em H} band. 
The resolved components together account for at least 25\% of the flux at {\em K}$'$
and 35\% of the flux at {\em H}.
The smallest resolved structure may be related to the radius of dust
sublimation at the inner part of a disk or envelope heated directly by
stellar radiation. However, the actual size of this region (8~mas or 5~AU), is about
six times too large for the supposed luminosity of V376~Cas. If
V376~Cas possesses a disk, it is likely to be viewed close to edge-on,
and this would lead to the photometrically determined luminosity being
an underestimate of the true luminosity. Furthermore, the outer part
of any disk probably obscures the inner disk edge, where the dust
sublimation region would be situated. The NIR flux observed would then
most likely arise from scattering from the inner edge of a cavity
above the disk plane. This hypothesis is further strengthened by the
high degree of linear polarization displayed by this object.

\begin{acknowledgements}
We thank John Monnier for providing us with data used in
Fig.~\ref{radiuslum} and Sasha Men'shchikov and Thomas Driebe for
helpful discussions. We also thank the referee, Peter Tuthill,
whose comments have led to significant improvements in the 
manuscript.
\end{acknowledgements}

\end{document}